\documentclass[aps,prd,preprint,superscriptaddress]{revtex4}
\usepackage{graphicx}
\usepackage{amsmath}
\usepackage{bm}% bold math
\usepackage{color}% bold math

\usepackage{soul}

\def\ee{\end{eqnarray}}

\def\=:{=\hspace{-.7em}\raisebox{1.1ex}{.}\hspace{.1em}\raisebox{-0.2ex}{.} }
\def\ee{\end{eqnarray}}

\def\=:{=\hspace{-.7em}\raisebox{1.1ex}{.}\hspace{.1em}\raisebox{-0.2ex}{.} }

\newcommand {\beq}{\begin{eqnarray}}
\newcommand {\eeq}{\end{eqnarray}}
\newcommand {\non}{\nonumber\\}

\newcommand {\1}[1]{\frac{1}{#1}}

\newcommand {\del}{\partial}

\begin{document}

%\preprint{}

\title{Fractional vortex molecules and 
vortex polygons\\ in a baby Skyrme model
}


\author{Michikazu Kobayashi}
\affiliation{
Department of Basic Science, University of Tokyo, Komaba 3-8-1, Meguro-ku, Tokyo 153-8902, Japan
}
\affiliation{
Department of Physics, Kyoto University, Oiwake-cho, Kitashirakawa, Sakyo-ku, Kyoto 606-8502, Japan
}
\author{Muneto Nitta}
\affiliation{Department of Physics, and Research and Education Center for Natural 
Sciences, Keio University, Hiyoshi 4-1-1, Yokohama, Kanagawa 223-8521, Japan
}
%\thanks{}



\date{\today}
\begin{abstract}
We construct a molecule of fractional vortices with fractional topological lump charges 
as a baby Skyrmion with the unit topological lump charge  
in the anti-ferromagnetic (or XY) baby Skyrme model, 
that is, an $O(3)$ sigma model with a four derivative term and 
an anti-ferromagnetic or XY-type potential term
quadratic in fields. 
We further construct configurations with 
topological lump charges $Q \leq 7$ and 
find that bound states of vortex molecules constitute 
regular polygons with $2Q$ vertices as vortices,  
where the rotational symmetry $SO(2)$ in real space  
is spontaneously broken into 
a discrete subgroup ${\bf Z}_Q$.
We also find metastable and arrayed bound states of 
 fractional vortices for $Q=5,6$. 
On the other hand, we find 
for $Q=7$ that the regular polygon 
is metastable and the arrayed bound state is stable.
We calculate binding energies of all configurations.

\end{abstract}
%\keywords{}

\maketitle

\section{Introduction}

Vortices are topological solitons present in various physical systems from field theory \cite{Manton:2004tk}
 and cosmological models \cite{Vilenkin:2000} 
to condensed matter systems \cite{Volovik:2003}.  
In particular, they play essential roles in condensed matter systems 
such as superconductors, superfluids, 
magnetism, quantum Hall states, 
nematic liquids, optics, and so on. 
One of the exotic aspects common in condensed matter 
systems but not familiar in high energy physics and cosmology 
are vortex molecules, which have been studied in 
multi-component Bose-Einstein condensates (BECs) \cite{Son:2001td,Kasamatsu:2004,Kasamatsu:2005,
Eto:2012rc,Cipriani:2013nya,Cipriani:2013wia,Eto:2013},
multi-gap superconductors 
\cite{Babaev:2002,Goryo:2007,Nitta:2010yf}, 
superfluid $^3$He (as a double core vortex) \cite{Volovik:2003}, 
and nonlinear optics \cite{optics}. 
In the cases of BECs \cite{Son:2001td} and 
superconductors \cite{Tanaka:2001}, 
fractional vortices in two different components 
with fractional topological charges constitute 
a meson-like bound state 
with the unit topological charge in total. 
However, a crucial difference between these two systems  
is that 
a repulsion between vortices is exponentially suppressed 
in superconductors 
due to the Higgs mechanism in the presence of a gauge field  
while a repulsion between vortices is polynomially reduced 
in BECs \cite{Eto:2011wp} 
in the absence of a gauge field. 
Consequently, 
vortex molecules are stable and visible in BECs  
because of a balance between the vortex repulsion 
and the domain wall tension,
while in superconductors they can be seen, in principle, 
only at temperatures above a certain critical temperature 
by a mechanism similar to 
the Berezinskii-Kosterlitz-Thouless transition \cite{Goryo:2007,Nitta:2010yf}.
However, stable vortex molecules in BECs are 
global vortices winding around a global $U(1)$ symmetry,
and consequently their energies are  
logarithmically divergent with respect to the system size;  
they are infinitely heavy in infinite space, 
and thereby they are not very realistic 
in high energy physics or cosmology. 

In this paper, we propose a field theoretical model 
admitting a vortex molecule with {\it finite} energy, 
motivated by these condensed matter systems.
We consider an $O(3)$ nonlinear sigma model 
on the target space $S^2$ in $d=2+1$ dimensions, 
described by a unit three-vector of scalar fields 
${\bf n}(x)=(n_1(x),n_2(x),n_3(x))$  
with the constraint ${\bf n}^2=1$, 
which is equivalent to a ${\bf C}P^1$ model.
The $O(3)$ model admits 
lumps or sigma model instantons 
characterized by $\pi_2(S^2) \simeq {\bf Z}$ \cite{Polyakov:1975yp} 
as a relative of vortices. 
We consider a potential term motivated by 
condensed matter systems admitting vortex molecules.
The potential terms make lumps unstable against shrinkage, in general, 
as can be inferred from the scaling argument \cite{Derrick:1964ww}, 
so we also consider a four derivative (Skyrme) term, 
by which the lumps are stabilized 
to become baby Skyrmions \cite{Piette:1994ug}.  
In the context of baby Skyrmions, 
the potential terms of the type 
$V = m^2 (1-n_3)$ \cite{Piette:1994ug} 
or of the type $V = m^2 (1-n_3^2)$ 
\cite{Kudryavtsev:1997nw,Weidig:1998ii,Harland:2007pb,Nitta:2012kk} 
have already been studied. 
The latter admits two discrete vacua $n_3=\pm 1$ and 
a domain wall interpolating between them 
\cite{Abraham:1992vb,
Kudryavtsev:1997nw,Harland:2007pb,Nitta:2012kj}, 
and a baby Skyrmion 
is in the shape of a twisted closed domain wall \cite{Weidig:1998ii,Kobayashi:2013ju}.
In our previous papers \cite{Nitta:2012xq,Nitta:2012wi,Kobayashi:2013ju}, 
we considered both types of potential terms 
$V = m^2 (1-n_3^2) + \beta^2 n_1$ in the regime $\beta \ll m$.  
In this case, a straight domain wall can 
absorb lumps as sine-Gordon kinks  
\cite{Nitta:2012xq,Nitta:2012wi,Kobayashi:2013ju}, 
and a baby Skyrmion is in the form of a domain wall ring 
with a sine-Gordon kink \cite{Kobayashi:2013ju}. 
In condensed matter physics, 
the quadratic potential admitting two vacua 
is known as the Ising-type in ferromagnets, 
so we may refer to this model as 
the Ising (or ferromagnetic) baby Skyrme model. 

%%%%%%%%%%%%%%%%%%%%%%
\begin{figure}[h]
\begin{center}
\vspace{-5cm}
\includegraphics[width=0.6\linewidth,keepaspectratio]{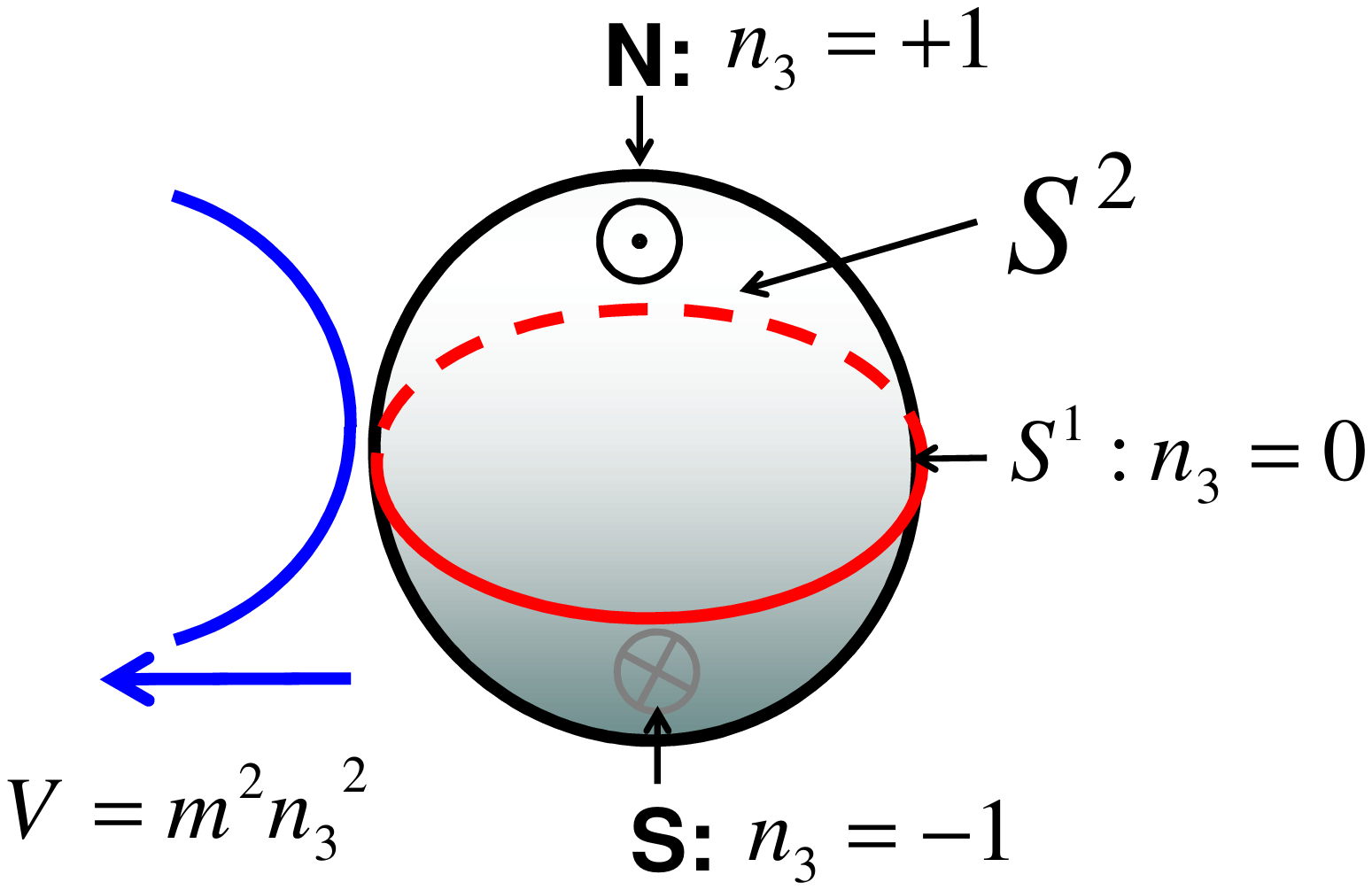}
\hspace{-3cm}
\includegraphics[width=0.5\linewidth,keepaspectratio]{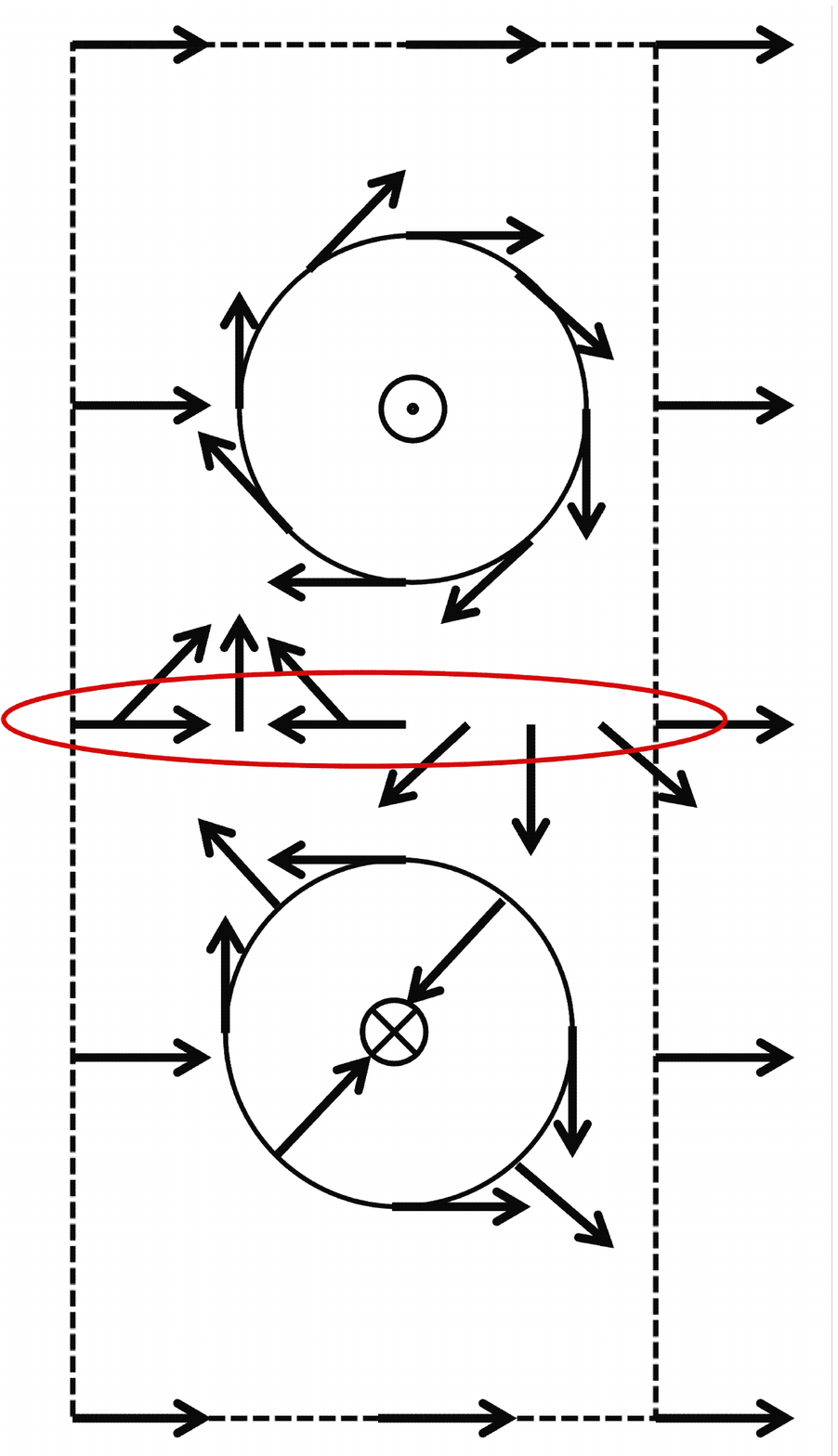}
\hspace{-3cm}
\caption{(a) The target space and the potential. 
(b) A vortex molecule as one lump. 
Here, $\leftarrow$, $\downarrow$, $\rightarrow$ and 
$\uparrow$ represent ${\bf n} =(-1,0,0), (0,-1,0), 
(1,0,0)$ and $(0,1,0)$, respectively. 
The north and south poles ${\bf n}=(0,0,1)$ and $(0,0,-1)$ 
are denoted by $\odot$ and $\otimes$, respectively. 
\label{fig:ns-vortices}
}
\end{center}
\end{figure}
%%%%%%%%%%%%%%%%%%%%%%
Here, we consider a potential 
of the XY-type or of anti-ferromagnets, 
$V =  m^2 n_3^2$ \cite{Jaykka:2010bq}. 
The model can be referred as
the XY (or anti-ferromagnetic) baby Skyrme model.
The vacua characterized by $n_3=0$ 
are $S^1$ at the equator of the target space $S^2$, 
as in Fig.~\ref{fig:ns-vortices} (a). 
One lump solution is schematically drawn 
in Fig.~\ref{fig:ns-vortices} (b), 
where we chose ${\bf n}=(1,0,0)$ as the vacuum 
at the boundary.
One can find two separated half lumps (merons)
whose centers are mapped to 
the north and south poles ${\bf n}=(0,0,\pm 1)$ 
of the target space $S^2$. 
These half lumps
are separated because the vacua $S^1$ appear 
between them  
as indicated by an ellipse in Fig.~\ref{fig:ns-vortices} (b).
Therefore, a lump is non-axisymmetric, 
unlike those for the massless case and the Ising-type case with $\beta = 0$ in which lumps are axisymmetric. 
There, the vacuum $U(1)$ winds (counter)clockwise 
between the two half lumps 
so that these half lumps are (anti-)global vortices. 
While isolated (anti-)global vortices have 
logarithmically divergent energy 
in the infinite system size, a pair of global and anti-global 
vortices have finite energy. 
They attract each other and collapse in the absence of 
the Skyrme term, while the Skyrme term forms a vortex molecule.  
We numerically construct fractional vortex molecules
with $1 \leq Q \leq 7$ 
by a relaxation method.   
The configuration of $Q=1$, in fact, looks what we expected. 
For $Q=2$, we find that 
two molecules face to each other 
with opposite orientations to constitute a regular square.  
Since two kinds of constituent vortices are placed 
at the vertices, the configuration is ${\bf Z}_2$ symmetric.  
For $Q=3$, we find a regular hexagonal 
 structure of vortex molecules like a benzene. 
This is ${\bf Z}_3$ symmetric. 
The configurations with $Q=2,3$ resemble 
those recently found in a two-component BEC under rotation 
\cite{Cipriani:2013nya}. 
Furthermore, for $Q=4,5,6$, we find regular octagonal, decagonal, and dodecagonal structures of fractional vortices 
with ${\bf Z}_4$, ${\bf Z}_5$, ${\bf Z}_6$ symmetries.
Therefore, in general, we expect, for the topological charge $Q$, 
a polygon with $2Q$ fractional vortices at vertices
with ${\bf Z}_Q$ symmetry \cite{Jaykka:2010bq}. 
We also find metastable and arrayed bound states of 
 fractional vortices for $Q=5,6$. 
These configurations are obtained by squeezing 
the corresponding stable polygons, 
and they have slightly higher energies.
We also find that the regular polygon 
is metastable and the arrayed bound state is stable 
for $Q=7$. 
Finally, we calculate binding energies of all the configurations.

Conventional lumps in the massless $O(3)$ sigma model are 
invariant under a combination of 
the $U(1)$ rotation of the target space 
and the space rotation. 
The lumps spontaneously break the other linear combination of 
the two $U(1)$ symmetries. 
On the other hand, 
our solutions spontaneously break the purely space rotation.
Note that non-axisymmetric configurations were known 
before for higher topological charges 
in the baby Skyrme model \cite{Delsate:2012hz}, 
while even one baby Skyrmion is non-axisymmetric in our model. 
In this regard, our solutions are similar to those in 
Ref.~\cite{Jaykka:2011ic}, 
in which non-axisymmetric molecular configurations 
were found in the model with a more complicated potential.
We would like to emphasize that 
our potential is quite common 
in anti-ferromagnets and two-component BECs 
and that it is natural.

A closely related model is
a $U(1)$ gauged supersymmetric ${\bf C}P^1$ model 
where a rotation along the $n_3$ axis is gauged 
\cite{Bagger:1982fn}. 
The gauge symmetry induces 
the potential $V =  m^2 n_3^2$ known 
as the D-term potential 
where supersymmetry requires $m$ to coincide with 
the gauge coupling $e$. 
In this model, a lump is decomposed into 
two fractional gauged 
Bogomol'nyi-Prasad-Sommerfield (BPS) vortices, 
and these two vortices can be placed with arbitrary separation 
because no force is present between BPS vortices 
\cite{Schroers:1995he,Nitta:2011um}. 
Each constituent vortex carries half of 
the lump charge characterized by 
$\pi_2(S^2)$, as in our model. 
A set of fractional vortices carrying the total unit charge 
was also found in supersymmetric gauge theories 
and sigma models \cite{Eto:2009bz}.

This paper is organized as follows. 
After our model is explained in Sec.~\ref{sec:model},
we give numerical solutions of vortex molecules 
in Sec.~\ref{sec:molecule}. 
Section \ref{sec:summary} is devoted to a summary 
and discussion.

%%%%%%%%%%%%%%%%%%%%%%%%%%
\section{The model \label{sec:model}}
We consider an $O(3)$ sigma model in $d=2+1$ dimensions
described by a three vector of scalar fields 
${\bf n} (x)= (n_1(x),n_2(x),n_3(x))$  
with a constraint ${\bf n} \cdot {\bf n} = 1$.
The Lagrangian of our model is given by
\beq
&& {\cal L} = \1{2} \del_{\mu}{\bf n}\cdot \del^{\mu} {\bf n} 
 - {\cal L}_4({\bf n})
 - V({\bf n}), \quad  
 \label{eq:Lagrangian}
\eeq
with $\mu=0,1,2$. 
Here, the four derivative (baby Skyrme) term is expressed as
\beq
{\cal L}_4 ({\bf n})
= \kappa F_{\mu\nu}^2 
= \kappa  \left[{\bf n} \cdot 
 (\partial_{\mu} {\bf n} \times \partial_{\nu} {\bf n} )\right]^2
= \kappa (\partial_{\mu} {\bf n} \times \partial_{\nu} {\bf n} )^2 ,
\quad F_{\mu\nu} \equiv {\bf n} \cdot 
 (\partial_{\mu} {\bf n} \times \partial_{\nu} {\bf n} ).
\eeq
In this paper, we take the potential term to be \cite{Jaykka:2010bq} 
\beq
V({\bf n}) = m^2 n_3^2 . 
\label{eq:pot}
\eeq 
This potential is known in anti-ferromagnets 
and the XY model, 
while the potential in the form of ferromagnets 
$m^2 (1-n_3^2)$
was studied before in 
Refs.~\cite{Weidig:1998ii,Kudryavtsev:1997nw,Harland:2007pb,Nitta:2012kk}. 
The energy density of static configurations is
\beq
 \mathcal{E} \equiv \1{2} ( \del_a{\bf n}\cdot \del^a {\bf n} )  
+ {\cal L}_4({\bf n}) + V({\bf n}), \quad (a=1,2).
\eeq

Itroducing the projective coordinate 
$u (\in {\bf C})$ of ${\bf C}P^1$ 
by
\beq
 n_i = \phi^\dagger \sigma_i \phi, \quad
 \phi^T = (1, u)/\sqrt{1+|u|^2},
\label{eq:projective-coordinate}
\eeq
the Lagrangian (\ref{eq:Lagrangian}) 
can be rewritten in the form of the ${\bf C}P^1$ model 
with potential terms, given by
\beq
&& {\cal L} = 
2\frac{\partial_{\mu} u^* \partial^{\mu} u}
  {(1 + |u|^2)^2} 
- 8 \kappa \frac{(\del_{\mu}u^* \del^{\mu}u)^2 
 - |\del_{\mu} u \del^{\mu} u|^2}
 {(1+|u|^2)^4} 
 - V \label{eq:CP1} \\ 
&& V= m^2 D_3^2,  
\quad 
 D_3 \equiv  \frac{1 - |u|^2}{1 + |u|^2} = n_3,
\eeq
Here, $g_{uu^*}=1/(1+|u|^2)^2$ is the K\"ahler (Fubini-Study) metric 
of  ${\bf C}P^1$,  
$g^{uu^*}=(1+|u|^2)^2$ is its inverse, 
and 
$D_i = n_i$ are the moment maps (or the Killing potentials) 
of the $SU(2)$ isometry 
generated by $\sigma_i$.
If we gauge  the isometry generated by the generator 
$\sigma_3$ with gauge coupling $e$,  
the potential $V=e^2 D_3^2$ (with $m=e$) is known as 
the D-term potential in 
the supersymmetric 
$U(1)$ gauged ${\bf C}P^1$ model \cite{Bagger:1982fn}.

%%%%%%%%%%%%%%%%%%%%%%%%%%%%%%%%%%%%%%%%%%%%%%%%%%
\section{Vortex molecules \label{sec:molecule}}

The topological charge of the lump $\pi_2(S^2) \simeq {\bf Z}$ 
is given by
\beq
 Q 
&=& 
 \1{4 \pi} \int d^2x\:  F_{12}
= \1{4 \pi}  \int d^2x\: {\bf n}\cdot 
(\partial_1 {\bf n} \times \partial_2 {\bf n} )
=
 \1{4 \pi}  \int d^2x\:
\epsilon_{ijk}
n_i \partial_{1} n_j  \partial_{2} n_k  \non \label{eq:lump-charge}
&=& \1{2 \pi}  \int d^2x 
{i(\del_1 u^* \del_2 u - \del_2 u^* \del_1 u)
\over (1+|u|^2)^2} .
\eeq
In the presence of the potential, 
a lump is unstable to shrinking from 
the Derick's scaling argument \cite{Derrick:1964ww}. 
It can be stabilized in the presence of the baby Skyrme term, 
resulting in a baby Skyrmion.  
%%%%%%%%%%%%%%%%

%%%%%%%%%%%%%%%%%%%%%%
\begin{figure}%[htb]
\begin{center}
\vspace{-6cm}
\includegraphics[width=0.97\linewidth,keepaspectratio]{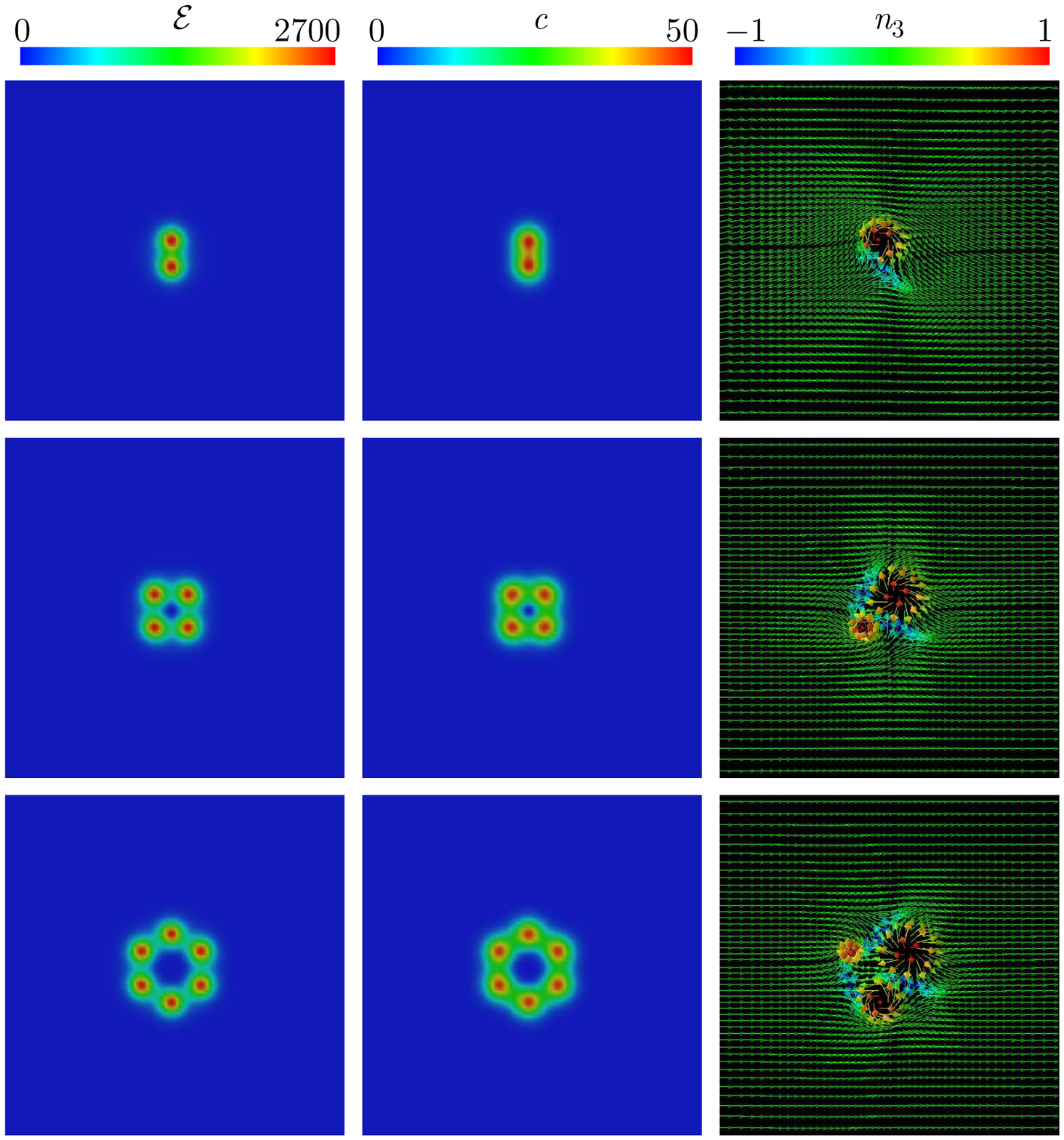}
\caption{Bound states of fractional vortex molecules in the region 
$-1.16 < x_a < 1.16$.
The topological charges are $Q=1,2,3$ from top to bottom.
The left and middle panels represent the total energy densities $\mathcal{E}(x)$ and the topological charge densities 
$c(x) \equiv F_{12}/4\pi$, respectively.
In the right panels, the arrows denote the scalar fields of 
the three-vector ${\bf n}(x)$,
where color represents the value of $n_3$ from which one can find 
whether the constituent fractional vortices are of N or S.
\label{fig:molecule}}
\end{center}
\end{figure}
%%%%%%%%%%%%%%%%%%%%%%
%%%%%%%%%%%%%%%%%%%%%%
\begin{figure}%[b]%[htb]
\begin{center}
\vspace{-6cm}
\includegraphics[width=0.97\linewidth,keepaspectratio]{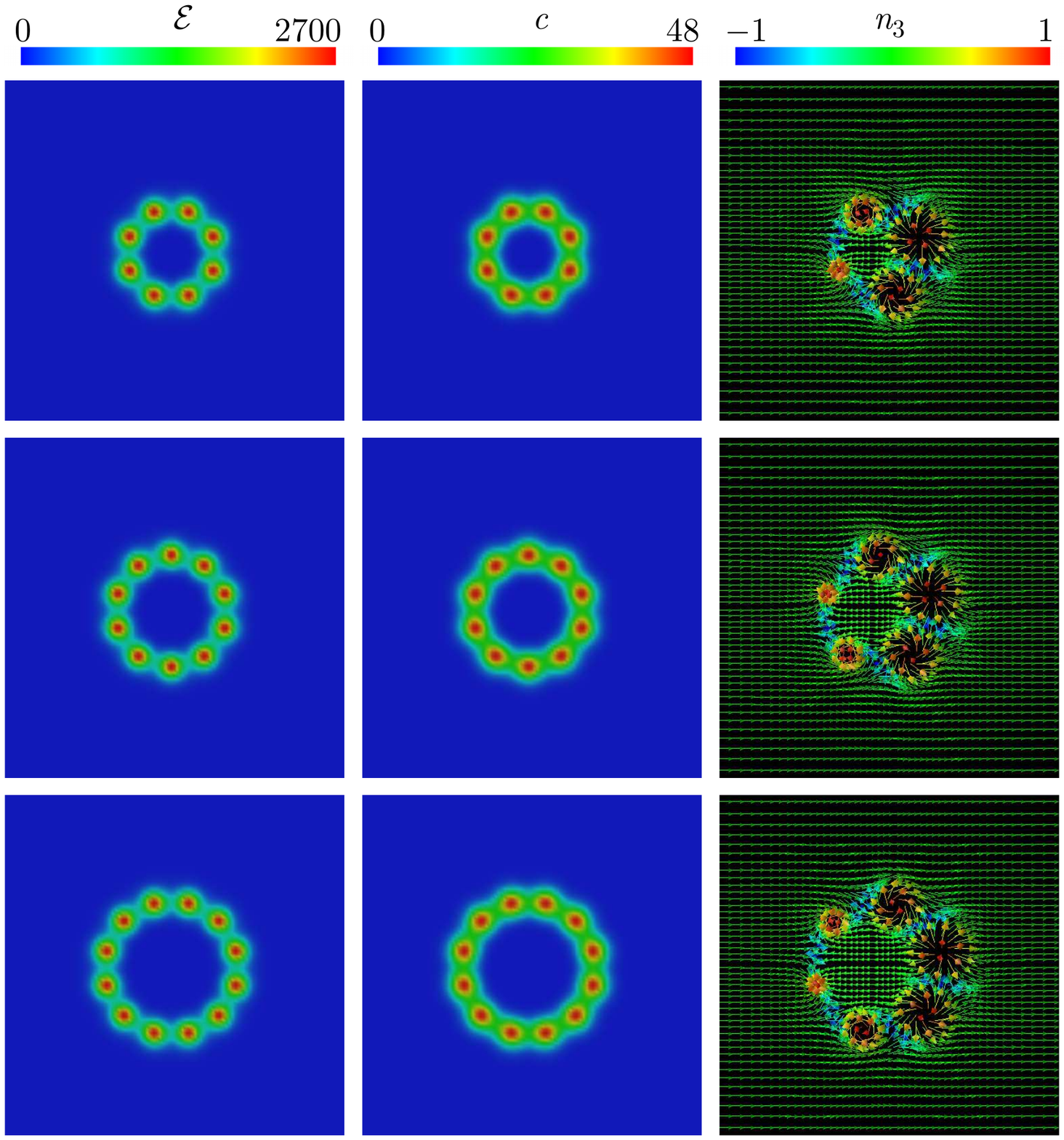}
\caption{Bound states of fractional vortex molecules in the region $-1.16 < x_a < 1.16$.
The topological charges are $Q=4,5,6$ from top to bottom.
Details of the plots are the same as those shown in Fig.~\ref{fig:molecule}.
\label{fig:molecule-2}}
\end{center}
\end{figure}
%%%%%%%%%%%%%%%%%%%%%%
%%%%%%%%%%%%%%%%%%%%%%
\begin{figure}%[b]%[htb]
\begin{center}
\vspace{-10cm}
\includegraphics[width=0.97\linewidth,keepaspectratio]{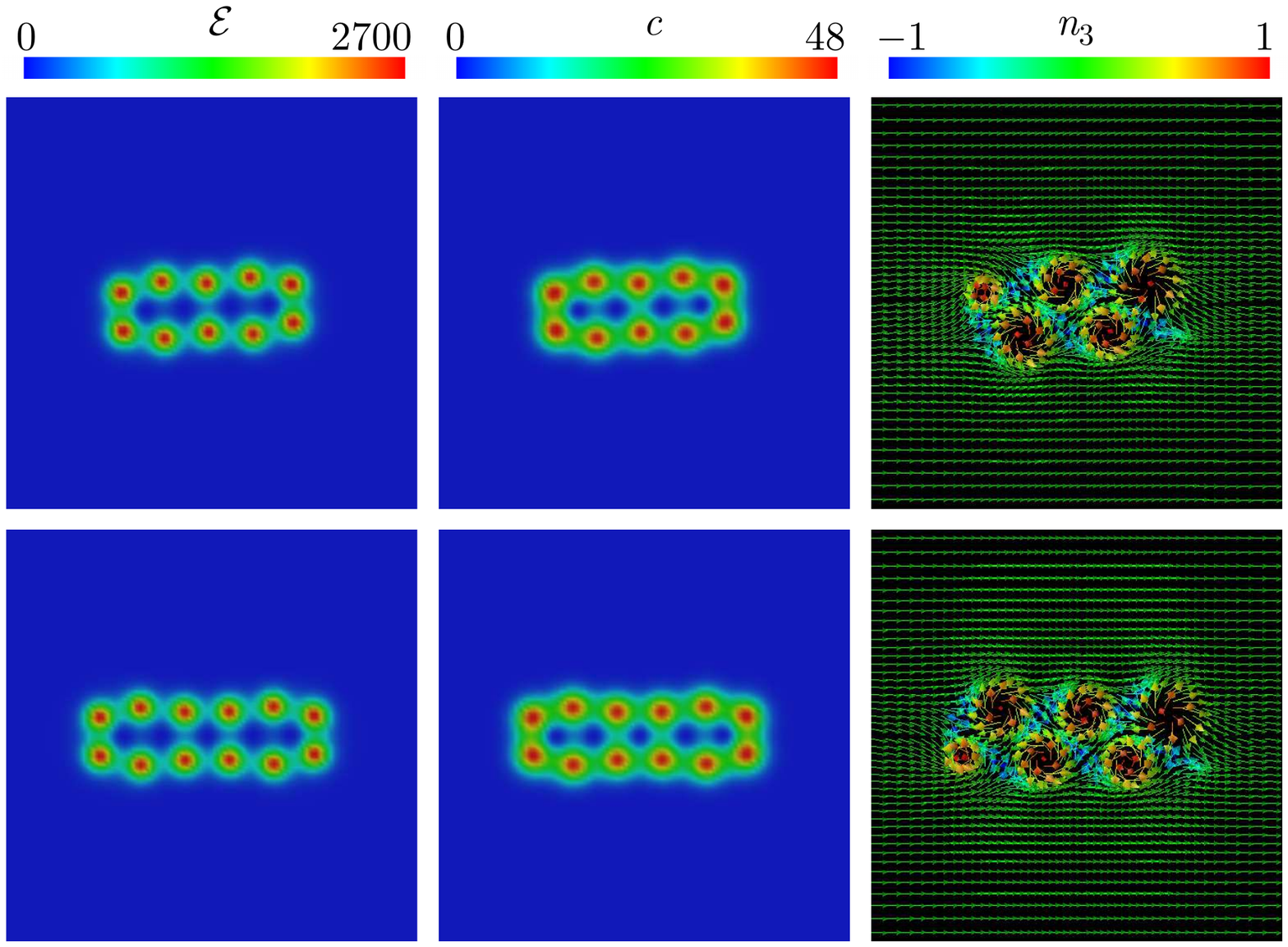}
\caption{Metastable and arrayed  bound states of fractional vortex molecules 
in the region $-1.16 < x_a < 1.16$.
The topological charges are $Q=5,6$ for the top and bottom.
Details of the plots are the same as those shown in Fig.~\ref{fig:molecule}.
\label{fig:molecule-3}}
\end{center}
\end{figure}
%%%%%%%%%%%%%%%%%%%%%%
%%%%%%%%%%%%%%%%%%%%%%
\begin{figure}%[b]%[htb]
\begin{center}
\vspace{-3cm}
\includegraphics[width=0.97\linewidth,keepaspectratio]{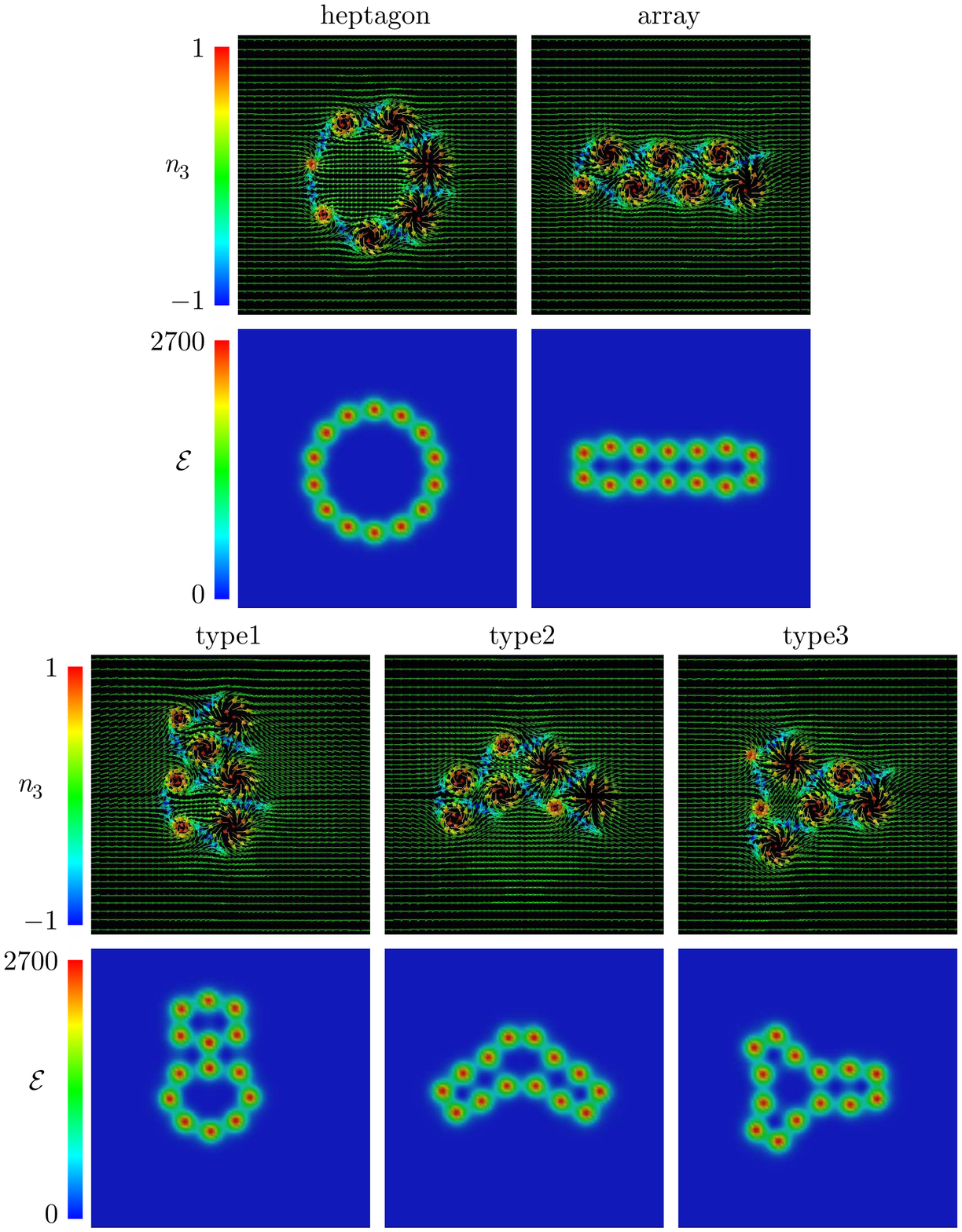}
\caption{Stable (arrayed) and metastable (heptagon and the other 3 types of) bound states of
fractional vortex molecules 
for the topological charge $Q=7$.
The top and bottom panels of each configuration represent the three-vector ${\bf n}(x)$ and the total energy density $\mathcal{E}(x)$, 
respectively,  
in the region $-1.16 < x_a < 1.16$.
\label{fig:molecule-4}}
\end{center}
\end{figure}
%%%%%%%%%%%%%%%%%%%%%%

We construct numerical solutions of 
fractional vortex molecules with the topological lump charge 
$1 \leq Q \leq 7$.
As the numerical parameters, we fix $\kappa = 0.002$ and $m^2 = 800$.
We obtain the stationary state using the relaxation method: introducing the parameter $\tau$ and the $\tau$-dependence of $n_i(\tau)$, 
and finding the asymptotic solution $n_i(\tau \to \infty)$ of the equation
\begin{align}
\frac{\partial n_i}{\partial \tau} = - \frac{\delta \mathcal E}{\delta n_i}, \label{eq-relaxation}
\end{align}
under the constraint $n_1^2 + n_2^2 + n_3^2 = 1$.
The detailed numerical procedure is shown in the Appendix \ref{appendix-numeric}.
As the initial state $n_i(\tau = 0)$, we give the ansatz for $\kappa = m = 0$:
\begin{align}
n_1 = \cos f(r), \quad
n_2 = - \cos (Q \theta) \sin f(r), \quad
n_3 = \sin (Q \theta) \sin f(r),
\label{eq:polygon-ansatz}
\end{align}
with the monotonically decreasing function satisfying
\begin{align}
f(r \to \infty) \to 0, \quad
f(r \to 0) \to \pi.
\end{align}
The topological charge $Q$ given by Eq. \eqref{eq:lump-charge} is invariant for arbitrary $\tau$.

Our stable solutions are given in Fig.~\ref{fig:molecule} for $Q = 1,2,3$ and Fig.~\ref{fig:molecule-2} for $Q = 4,5,6$. 
For the unit topological charge, $Q=1$,  
one can find a fractional vortex molecule as we expected; 
two fractional vortices oppositely 
wind around the equator $S^1$ of the target space $S^2$, 
and their cores are filled by 
the north $n_3=+1$ and south  $n_3=-1$ poles of $S^2$, 
as can be seen in the plot of $n_3$ in Fig.~\ref{fig:molecule}. 
The vacua $S^1$ appear between the two half lumps 
separating them, 
as can be seen from both the plots of the 
energy density and topological charge density. 
Although all of them are (anti-)global vortices 
having logarithmically divergent energy, 
a pair them has finite energy. 
They attract each other, but the Skyrme term prevents 
the collapse. 

For $Q=2$, 
two vortex molecules face to each other 
with opposite orientations. 
Since molecules attract each other with these orientations, 
they make a bound state to constitute a square. 
Since the same vortices N or S are placed at each pair 
of diagonal corners, 
the configuration is ${\bf Z}_2$ axisymmetric.

For $Q=3$, 
three vortex molecules with six fractional vortices 
constitute a hexagon 
with a  ${\bf Z}_3$ axisymmetry. 
These structures of $Q=2,3$ resemble 
those in a vortex lattice recently found in 
a two-component BEC under rotation 
\cite{Cipriani:2013nya}.

For $Q=4,5,6$, the situation is almost the same; {\it i.e.},
one finds that four, five, and six vortex molecules 
with eight, ten, and eleven fractional vortices constitute
an octagon, decagon, and dodecagon
with ${\bf Z}_4$, ${\bf Z}_5$, ${\bf Z}_6$ axisymmetries, respectively.

In general, for the topological charge $Q$, 
we expect 
$2Q$ fractional vortices to be placed on a circle 
in a ${\bf Z}_Q$ axisymmetric way.  
In all the cases, 
one can see that the topological lump charge density 
 is distributed around the fractional vortices, 
and each of them carries a half lump charge.
The rotational symmetry $SO(2)$ in the $x^1$-$x^2$ plane 
is spontaneously broken in all cases to 
a discrete subgroup ${\bf Z}_Q$.

To investigate the stability of vortex polygons, 
we also choose a randomly placed lump solution
\begin{align}
u = \sum_{i = 1}^Q \frac{e^{i \varphi_i}}{x_1 - x_{0i} + i x_2 - y_{0i}}, \label{eq:random-molecule}
\end{align}
as initial states $n_i(\tau = 0)$ for the relaxation.
Here, the projective coordinate $u$ is defined in 
Eq.~\eqref{eq:projective-coordinate}, and $x_{0i}$, $y_{0i}$, and $\varphi_i$ are the real random numbers.
The initial state \eqref{eq:random-molecule} indicates that $Q$ single vortex molecules are placed at $(x_1, x_2) = (x_{0i}, y_{0i})$ with the angle $\varphi_i$, having the topological charge $Q$.

For $Q = 1,2,3,4$, any initial state of Eq.~\eqref{eq:random-molecule} relaxes to vortex polygons as shown in Figs. \ref{fig:molecule} and \ref{fig:molecule-2}, 
while, for $Q = 5, 6$, several initial states relax to  
metastable states that are different from 
the vortex decagon and dodecagon, {\it i.e.}, vortex array states 
as in Fig.~\ref{fig:molecule-3}, 
the energy of which is larger than vortex polygon states.

We also find 
five (meta)stable bound states 
for $Q=7$ as in Fig.~5. 
Unlike the cases with $Q<7$, 
the arrayed bound state is at the absolute minimum.
The regular polygon and 
the other three bound state are metastable 
at local minima.

%%%%%%%%%%%%%%%%%%%%%%%
\begin{table}
\begin{center}
\begin{tabular}{c||c|c|c|c|c}
$Q$ & $E_{\mathrm{grad}}$ & $E_{\mathrm{sk}}$ & $E_{\mathrm{pot}}$ & $E$ & $E_{\mathrm{bind}}$ \\ \hline\hline
1 & 29.04 & 8.840 & 8.829 & 46.71 & \\ \hline
2 & 53.35 & 16.96 & 16.96 & 87.26 & 6.153 \\ \hline
3 & 78.45 & 25.17 & 25.17 & 128.8 & 11.33 \\ \hline
4 & 103.8 & 33.44 & 33.44 & 170.7 & 16.14 \\ \hline
5 (decagon) & 129.3 & 41.73 & 41.73 & 212.7 & 20.79 \\
5 (array) & 129.8 & 41.68 & 41.68 & 213.1 & 20.42 \\ \hline
6 (dodecagon) & 154.8 & 50.03 & 50.03 & 254.9 & 25.35 \\
6 (array) & 155.2 & 49.90 & 49.90 & 255.0 & 25.22 \\ \hline
7 (heptagon) & 180.4 & 58.34 & 58.34 & 297.1 & 29.86 \\
7 (array) & 180.7 & 58.14 & 58.13 & 297.0 & 29.97 \\
7 (type 1) & 181.3 & 58.29 & 58.28 & 297.9 & 29.09 \\
7 (type 2) & 181.2 & 58.22 & 58.22 & 297.6 & 29.31 \\
7 (type 3) & 181.6 & 58.32 & 58.32 & 298.3 & 28.68
\end{tabular}
\caption{Gradient energy $E_{\mathrm{grad}}$, Skyrme energy $E_{\mathrm{sk}}$, potential energy $E_{\mathrm{pot}}$, total energy $E$, and binding energy $E_{\mathrm{bind}}$ for vortex molecules.
\label{table:molecule}}
\end{center}
\end{table}
%%%%%%%%%%%%%%%%%%%%%%%%%%%%%%%%%%%%%%%%%%%%%%%%%%%%%%%%%%%%%%%%%%%%%%%
Finally, we show in Table \ref{table:molecule} 
the gradient energy $E_{\mathrm{grad}}$, the Skyrme energy $E_{\mathrm{sk}}$, the potential energy $E_{\mathrm{pot}}$, the total energy $E$, and the binding energy $E_{\mathrm{bind}}$ between vortex molecules, 
defined by
\begin{align}
\begin{array}{c}
\displaystyle
E_{\mathrm{grad}} = \frac{1}{2} \int d^2x\: \partial_a {\bf n} \cdot \partial^a {\bf n}, \quad
E_{\mathrm{sk}} = \int d^2x\: {\cal L}_4({\bf n}), \quad
E_{\mathrm{pot}} = \int d^2x\: V({\bf n}), \vspace{5pt} \\
\displaystyle
E = E_{\mathrm{grad}} + E_{\mathrm{sk}} + E_{\mathrm{pot}}, \quad
E_{\mathrm{bind}} = Q E(Q=1) - E(Q),
\end{array}
\end{align}
respectively. 
In our choice of numerical parameters, the gradient energy is dominant in the total energy.
Total energy deviates from the linear relation, {\it i.e.}, 
$E(Q) < Q E(Q = 1)$, and the
difference $E_{\mathrm{bind}}$ between $E(Q)$ and $Q E(Q = 1)$ corresponds to the binding energy between molecules,
which dominates in about $10$\% of the total energy.

%%%%%%%%%%%%%%%%%%%%%
\section{Summary and Discussion \label{sec:summary} }

We have constructed stable and metastable configurations of 
the fractional vortex molecules as lumps (baby Skyrmions) 
in the XY (or anti-ferromagnetic) baby Skyrme model 
which has the anti-ferromagnetic (XY) potential $V= m^2 n_3^2$.   
We have found that, for the unit charge $Q=1$, 
two fractional vortices whose centers are 
filled by the north and south poles of the 
target space are placed within a certain distance. 
We have found that, for $Q=2,3,4,5$ and $6$, bound states of 
two, three, four, five, and six vortex molecules 
constitute quadrangle, hexagonal, octagonal, decagonal, and dodecagonal 
vortex configurations, respectively.  
At least up to this topological number, 
${\bf Z}_Q$ symmetric vortex molecule configurations 
appear for the topological charge $Q$. 
Our configurations are all non-axisymmetric, and they 
spontaneously break the rotational symmetry in the $x$-$y$ plane.
While all vortex polygons are stable 
and at global minima, 
we have also found metastable and arrayed bound states of 
 fractional vortices for $Q=5,6$, 
which are obtained by squeezing 
the corresponding stable polygons 
and have slightly higher energies. 
We also find  for $Q=7$ that the arrayed bound state 
is at the absolute minimum and 
the regular polygon together with 
the other three bound state is metastable 
at a local minimum,  
unlike the cases with $Q<7$
Finally, we have calculated the binding energies of all the configurations.

As denoted in the Introduction, 
similar configurations of vortex molecules 
are present in condensed matter systems, such as 
two component BECs, 
described by two condensations (scalar fields) 
$\phi_1(x)$ and $\phi_2(x)$
with  the internal coherent (Josephson) 
coupling $\beta^2 \phi_1^*\phi_2 +$ c.c.
\cite{Son:2001td,Kasamatsu:2004,Kasamatsu:2005,
Eto:2012rc,Cipriani:2013nya}, 
where a four derivative term is not present. 
In these cases, 
the vortex molecules are global vortices, 
that is, they wind around a global $U(1)$ symmetry.   
If we gauge the common phase of the two components, 
they become semi-local vortices. 
If we send the scalar coupling and gauge coupling to infinity 
with keeping the internal coherent coupling, 
we obtain our model except for the Skyrme term 
where the Josephson coupling 
$\beta^2 \phi_1^*\phi_2 +$ c.c 
reduces to $\beta^2 n_1$, 
which we did not consider in this paper. 
We expect that semi-local vortices, 
 keeping the couplings, can make a stable 
vortex molecule if one properly adds a four derivative term.

In the ${\bf C}P^1$ model with a ferromagnetic potential, 
a Q-lump solution is known \cite{Leese:1991hr},
in which a $U(1)$ Nambu-Goldstone mode associated with 
the spontaneously broken $U(1)$ internal symmetry of the rotation 
in the $n_1$-$n_2$ plane in the target space 
is rotating in time. 
In our case, there is a Nambu-Goldstone mode associated with
the spontaneously broken rotational symmetry in real space,
instead of an internal $U(1)$ Nambu-Goldstone mode.
Consequently, we may have a Q-lump in our case as a spinning 
molecule.

%%%%%%%% 3D
If we promote our configuration linearly in $d=3+1$ dimensions,
it becomes a bound state of two cosmic strings.
As usual, 
the solution breaks translational symmetries in two transverse directions, resulting in two translational zero modes which propagate along the string. 
One added feature of our solutions is 
the existence of a Nambu-Goldstone mode associated with 
the spontaneously broken rotational symmetry 
along the string, resulting in a twisting wave propagating 
along the string. This is known as a ``twiston" in 
two-gap superconductors \cite{Tanaka:2007}. 
In cosmology,  our solutions  
can be regarded as some exotic cosmic strings 
with an internal structure.
For instance, 
it is an interesting question whether or not 
two such strings reconnect to each other when they collide.

We have found that, for the topological charge $Q$,
two kinds of vortices are placed at $2Q$ vertices 
of a regular polygon, 
on which a ${\bf Z}_Q$ symmetry acts.
In this regard, 
${\bf Z}_n$ symmetric vortex configurations 
were studied in the Abelian-Higgs model \cite{Arthur:1995eh}. 
This is equivalent to vortices on the orbifold ${\bf C}/{\bf Z}_n$ 
studied recently \cite{Kimura:2011wh}.  
Vortex polygons have also been studied in hydrodynamics 
for a long time \cite{fluid}.
Vortex polygons with less than seven vortices as vertices 
are shown to be stable, 
while those with more than seven are unstable. 
One example realized in nature is 
a vortex hexagon found by Cassini 
in the north poles of Saturn \cite{saturn}. 
In our case too, we have found 
vortex polygons up to 
$Q=6$ to be stable, which may be 
interesting compared with hydrodynamics.

%%%%%%%% N-components
In this paper, we have found a two vortex molecule, that is, 
a vortex dimer with the unit topological charge 
in the ${\bf C}P^1$ model with 
the anti-ferromagnetic potential. 
Stable three or $N$ vortex molecules, that is,   
vortex trimers or $N$-omers,  
are present in three or $N$ component BECs 
\cite{Eto:2012rc,Eto:2013}. 
Since two components $\phi_1,\phi_2$ with a constraint 
$|\phi_1|^2+|\phi_2|^2=1$ divided by $U(1)$  
imply the ${\bf C}P^1$ model, 
the same procedure for $N$ components 
yields a ${\bf C}P^{N-1}$ model with a certain potential term. 
While a ${\bf C}P^{N-1}$ generalization of 
the anti-ferromagnetic type potential 
was considered before \cite{AlvarezGaume:1983ab}  
and was found to admit parallel multiple domain walls \cite{Gauntlett:2000ib} 
or domain wall junctions or networks \cite{Eto:2005cp,Eto:2006pg},
depending on parameters in the potential,  
a ${\bf C}P^{N-1}$ generalization of 
the ferromagnetic (XY-type) of the potential 
has not been studied thus far in the ${\bf C}P^{N-1}$ model.
Skyrme-like terms in the ${\bf C}P^{N-1}$ model were studied before 
without \cite{Ferreira:2008nn,Liu:2009rz} 
and with \cite{Bergshoeff:1984wb,Eto:2012qda,Adam:2011hj}
supersymmetry.
A ${\bf C}P^2$ or ${\bf C}P^{N-1}$ generalization 
of the baby Skyrme model with 
an anti-ferromagnetic potential will 
admit vortex trimers or $N$-omers, respectively.

\section*{Note added}
After the paper was published, 
vortex polygons have been studied 
in a baby Skyrme model with 
a different potential \cite{Winyard:2013ada}.

\section*{Acknowledgements}

M.~N. thanks Minoru Eto for collaborations 
which motivated this work. 
This work is supported in part by 
Grant-in-Aid for Scientific Research 
(Grants No. 22740219 (M.K.) and No. 23740198 and No. 25400268 (M.N.)) 
and the work of M. N. is also supported in part by 
the ``Topological Quantum Phenomena'' 
Grant-in-Aid for Scientific Research 
on Innovative Areas (No. 23103515 and No. 25103720)  
from the Ministry of Education, Culture, Sports, Science and Technology 
(MEXT) of Japan. 

\begin{appendix}

\section{Detailed numerical procedure} \label{appendix-numeric}

Equation \eqref{eq-relaxation} can be solved as the steepest descent method:
\begin{align}
n_i(\tau + \Delta \tau) = n_i(\tau) - \Delta \tau \frac{\delta \mathcal{E}_\lambda(\tau)}{\delta n_i(\tau)},
\end{align}
where we omit the spatial dependence of $n_i$.
$\mathcal E_\lambda(\tau)$ is defined as
\begin{align}
\mathcal{E}_\lambda(\tau) = \mathcal{E}(\tau) - \lambda(\tau) \{n_1(\tau)^2 + n_2(\tau)^2 + n_2^2(\tau)\}.
\end{align}
Here, the Lagrange multiplier $\lambda(\tau)$ is fixed to satisfy $n_1(\tau + \Delta \tau)^2 + n_2(\tau + \Delta \tau)^2 + n_2(\tau + \Delta \tau)^2 = 1$.

For the space, to approximately consider the infinite space, we use the following scaling transformation:
\begin{align}
x_a = L \tanh^{-1} X_a
\end{align}
for $-1 < X_a < 1$, and consider the dependence of $n_i$ on $X_a$ instead of $x_a$,
where $L$ is the scaling parameter.
We use the square with the $(N+1)^2$ grid points.
On the $l$-th grid point in the $x_a$-direction, the value of $n_i(\{(l)_a\})$ is defined as
\begin{align}
n_i(\{(l)_a\}) \equiv n_i(\{ (\cos(l \pi / N))_a \}),
\end{align}
where $n_i(\{X_a\})$ is the value of $n_i$ at $\{X_a\} \equiv (X_1, X_2)$.
We omit the $\tau$-dependence on $n_i$ here.
For $l = 0$ or $N$, which corresponds to infinity, 
the value of $n_i(\{(l)_a\})$ is fixed to the ground state:
\begin{align}
\begin{split}
n_1(\{(0)_a\}) = n_1(\{(N)_a\}) = 1, \\
n_2(\{(0)_a\}) = n_2(\{(N)_a\}) = 0, \\
n_3(\{(0)_a\}) = n_3(\{(N)_a\}) = 0.
\end{split}
\end{align}
To calculate the spatial derivative of $n_i$, we use the spectral collocation method.
We expand $n_i$ in the Chebyshev polynomials:
\begin{align}
\tilde{n}_i(\{(j)_a\}) = \frac{1}{N} n_i(\{(0)_a\}) + \frac{2}{N} \sum_{l = 1}^{N-1} n_i(\{(l)_a\}) \cos\bigg(\frac{l j \pi}{N}\bigg) + \frac{(-1)^j}{N} n_i(\{(N)_a\}),
\label{eq:forward-Chebyshev}
\end{align}
where $\tilde{n}_i(\{(j)_a\})$ $(0 \leq j \leq N)$ is the $j$-th coefficient of the Chebyshev expansion in the $x_a$-direction.
The first and second spatial derivatives $\partial_a n_i$ and $\partial_a^2 n_i$ can be calculated as
\begin{align}
\begin{array}{c}
\displaystyle
\partial_a n_i = \frac{1 - X_a^2}{L} \frac{\partial n_i}{\partial X_a}, \quad
\partial_a^2 n_i = \frac{(1 - X_a^2)^2}{L^2} \frac{\partial^2 n_i}{\partial X_a^2} - \frac{2 X_a (1 - X_a^2)}{L^2} \frac{\partial n_i}{\partial X_a}, \vspace{5pt} \\
\displaystyle
\frac{\partial n_i(\{(l)_a\})}{\partial X_a} = \frac{1}{N} \tilde{m}_i(\{(0)_a\}) + \frac{2}{N} \sum_{j = 1}^{N-1} \tilde{m}_i(\{(j)_a\}) \cos\bigg(\frac{l j \pi}{N}\bigg) + \frac{(-1)^l}{N} \tilde{m}_i(\{(N)_a\}), \vspace{5pt} \\
\displaystyle
\frac{\partial^2 n_i(\{(l)_a\})}{\partial X_a^2} = \frac{1}{N} \tilde{s}_i(\{(0)_a\}) + \frac{2}{N} \sum_{j = 1}^{N-1} \tilde{s}_i(\{(j)_a\}) \cos\bigg(\frac{l j \pi}{N}\bigg) + \frac{(-1)^l}{N} \tilde{s}_i(\{(N)_a\}).
\label{eq:backward-Chebyshev}
\end{array}
\end{align}
Coefficients $\tilde{m}_i(\{(j)_a\})$, and $\tilde{s}_i(\{(j)_a\})$ satisfy the following recurrence relations:
\begin{align}
\begin{array}{c}
\displaystyle
\tilde{m}_i(\{(j)_a\}) = \tilde{m}_i(\{(j+2)_a\}) + 2 (j + 1) \tilde{n}_i(\{(j+1)_a\}), \quad \tilde{m}_i(\{(N)_a\}) = 0, \vspace{5pt} \\
\tilde{s}_i(\{(j)_a\}) = \tilde{s}_i(\{(j+2)_a\}) + 2 (j + 1) \tilde{m}_i(\{(j+1)_a\}), \quad \tilde{s}_i(\{(N)_a\}) = 0.
\end{array}
\end{align}
Equations \eqref{eq:forward-Chebyshev} and \eqref{eq:backward-Chebyshev} can be calculated by the fast Fourier transform algorithm.

In this paper, we fix $L = 0.5$, $N = 256$, $\Delta \tau = 10^{-6}$.

\end{appendix}

%%%%%%%%%%%%%%%%%%%%%%%%%%%%%%%%%%%%%%%%%%%%%%%%%%%%%%%%%%%%
%\newpage

%%%%%%%%%% References %%%%%%%%%%%%%%%%%%%%%%%%%
\newcommand{\J}[4]{{\sl #1} {\bf #2} (#3) #4}
\newcommand{\andJ}[3]{{\bf #1} (#2) #3}
\newcommand{\AP}{Ann.\ Phys.\ (N.Y.)}
\newcommand{\MPL}{Mod.\ Phys.\ Lett.}
\newcommand{\NP}{Nucl.\ Phys.}
\newcommand{\PL}{Phys.\ Lett.}
\newcommand{\PR}{ Phys.\ Rev.}
\newcommand{\PRL}{Phys.\ Rev.\ Lett.}
\newcommand{\PTP}{Prog.\ Theor.\ Phys.}
\newcommand{\hep}[1]{{\tt hep-th/{#1}}}
%%%%%%%%%%%%%%%%%%%%%%%%%%%%%%%%%%%%%%%%%%%%%%%

\end{document}